# Intelligent Adaptive Federated Byzantine Agreement for Robust Blockchain Consensus

**Erdhi Widyarto Nugroho[1], R.Rizal Isnanto[2], Luhur Bayuaji[3,4]**

[1] Doctoral Program of Information Systems, Diponegoro University Semarang, Indonesia
[2] Department of Computer Engineering, Faculty of Engineering, Diponegoro UniversitySemarang Indonesia
[3] Faculty of Data Science and Information Technology INTI International University Nilai, Malaysia
[4] Faculty of Information Technology, Universitas Budi Luhur Jakarta, Indonesia

*Keywords:*

Blockchain System
Consensus mechanisms
Intelegent Adaptive FBA
Quorum Slice
Fault Tolerance

**ABSTRACT**

The Federated Byzantine Agreement (FBA) achieves rapid consensus by relying on overlapping quorum slices. But this architecture leads to a high dependence on the availability of validators: when about one-fourth of validators go down, the classical FBA can lose liveness or fail to reach agreement. We thus come up with an Adaptive FBA architecture that can reconfigure quorum slices intelligently based on real-time validator reputation to overcome this drawback. Our model includes trust scores computed from EigenTrust and a sliding-window behavioral assessment to determine the reliability of validators.

We have built the Intelegent adaptive FBA model and conducted tests in a Stellar-based setting. Results of real-life experiments reveal that the system is stable enough to keep consensus when more than half of the validators (up to 62%) are disconnected, which is a great extension of the failure threshold of a classical FBA. A fallback mode allows the network to be functional with as few as three validators, thus showing a significant robustness enhancement. Besides, a comparative study with the existing consensus protocols shows that Adaptive FBA can be an excellent choice of the next generation blockchain systems, especially for constructing a resilient blockchain infrastructure.

*Corresponding Author:*

Erdhi Widyarto Nugroho
Department of Information Systems, Doctorate, Diponegoro University Semarang, Indonesia
Jl. Imam Bardjo No. 1-3 Semarang 50241
Email: erdhiwidyarto@students.undip.ac.id

## 1. INTRODUCTION

The release of Bitcoin in 2008 brought with it the term "blockchain" into public discourse, though it is unlikely that many at the time imagined how far the concept would travel. Originally intended merely as a means for recording digital monetary transactions without the need for a central authority, over time its core concept—a shared, tamper-evident data structure—found applicability in numerous other domains[1]. Once there was sufficient interest, different industries piloted blockchain for myriad reasons: some for transparency, some for decentralization, and most because it is very difficult to change information after it has been committed to storage [2]. It gradually crept out from being an odd piece of financial tooling toward becoming general resilient infrastructure [3].

The tech grew in steps. Old setups like Bitcoin were all about securing transactions. A major change arrived with Ethereum's smart contracts that allowed the rules of an app to operate directly on a network This opened up more uses for blockchain. Newer setups like Polkadot Cardano and Cosmos worked on problems of the past in growing systems that are different talking to other using parts based issues and many chain setups [4]. Despite the actions, some old concerns remain in terms of energy usage as it was busy and how to keep it fast when in open license free spaces [5].

Every blockchain system requires a consensus mechanism to function regardless of its design structure. The development of blockchain consensus mechanisms has led to the creation of Proof-of-Work (PoW) and Proof-of-Stake (PoS) and Practical Byzantine Fault Tolerance (PBFT) and Federated Byzantine Agreement (FBA) systems. The implementation of each consensus method brings advantages yet creates new system limitations. The security level of PoW is high but the system consumes large amounts of energy while operating at a slow pace [6]. The PoS and PoA systems enhance operational speed but they enable wealthy token owners and pre-approved validators to control network operations. [7] The PBFT system achieves quick finality but its ability to scale becomes limited because the number of messages required for consensus grows rapidly with each additional node [8].

FBA functions through a separate kind of operation. The consensus process takes place when different validators elect their trusted nodes for quorum slices because the nodes they chose coincide. The system architecture alters very slightly the time between the steps of the consensus process so that it remains decentralised while it does not allow participants to concentrate around rich validators [9]. The Stellar Consensus Protocol (SCP) gave a formal definition of FBA in 2015 before its implementation in Stellar and Ripple networks.FBA's equilibrium is still open to question, in fact, when it works under full conditions of the system [10].

The security of FBA is related to the extent to which the validators keep their trustworthiness. The system has the problem of agreement in the presence of failure of nodes and performing malicious activities since the stabilization of the quorum becomes difficult due to the failure of the overlap. The security of a Byzantine network is affected when the number of defective nodes reaches 25% of the network size. Various papers have identified numerous structural issues that have an impact on the system. One of the problems pinpointed in the work of Florian et al. (2022) was that certain quorum configurations become unstable when specific validators are unreachable [11]. García-Perez and Gotsman (2019) examined the security aspects of quorum to show that lack of overlap between nodes can lead to complete consensus failure [12]. Gaul and Liesen (2020) introduced the term "intactness" to refer to the measure of the quorums' resistance to failures in their network [13]. The ongoing research has mostly been done on static and small-scale networks which do not offer enough support for the development of adaptive models that extensively operate in the networks.

This research presents an Intelegent Adaptive FBA model that features the integration of repuration directly into the process of consensus. In the network each validator is measured using trust value by the EigenTrust and a sliding window. in the event of disruption in the network, the device slices adaptively regenerate to keep the network secure and the nodes connected.

We use simulation experiments based on the Stellar Consensus Protocol (SCP) framework to test our proposed design. The experiment was done by gradually increasing the number of validator failures until either the system losing its ability to reach consensus or the quorum slices losing their intersections. The data reveal that the adaptive configuration can sustain the FBA system up to the point where more than 50% of the validators have failed. Even with the fallback mode, the network can still function with a minimum of three active validators, thus, the Adaptive FBA system is proven to be resilient.

The adaptive configuration allows the FBA system to keep functioning even when more than half of its validators fail. Network isMaintained operational with only three validators even under the fallback mode, which is a clear indication of the high resilience level of the Adaptive FBA model. In the last chapter, the current work is compared with PoW, PoS, PoA, PBFT, and traditional FBA. In fact, the experimental findings show that the new system possesses a higher throughput, better fault tolerance, and enhanced scalability than the existing ones.

The major contributions of this work are the following three: A reputation-assessment method combining EigenTrust and sliding-window metrics for continuous validator evaluation;the regeneration model that Intelegent adaptively recreates the quorum-slice networks during failure condition; and A comparison framework demonstrating that the intelegent adaptive model brings forth scalability, resilience, and decentralization benefits that are superior to those of traditional models. In summation, the findings position FBA as a candidate for evolution into a trust-aware and self-recovering framework of consensus for blockchain networks in the future.

The paper is elaborated through five sections: Section II is dedicated to related work; Section III Experimental Design ; Section IV is about simulation results; and Section V is the wrap-up with design implications and directions for future scalable blockchain systems. The emphasis of this paper lies on the first three contributions:

(1) A reputation assessment method that combines the metrics of EigenTrust and sliding-window for continuous evaluation of the validators;

(2) a model for regeneration that Intelegent adaptively reconstructs quorum-slice networks under failure conditions; and

(3) a comparative framework that reveals the intelegent adaptive model's advantages in scalability, resilience, and decentralization.

Basically, the findings indicate that FBA may be transformed into a trust-aware and self-recovering consensus architecture for blockchain networks of the future.

## 2. RELATED WORK AND THEORETICAL BACKGROUND

### 2.1 State of the Art in FBA

The Federated Byzantine Agreement (FBA) has developed as an alternative consensus method for distributed systems. This is appreciated for its capacity to accommodate open and flexible network structures dependent on local trust relationships [14]. FBA has been implemented in several domains such as swarm robotics, smart grids, and electric-vehicle infrastructure [15], [16]. Some models like vFBAS and FBA-CPO show that they can perform well in situations without a leader and limited connectivity. They do so by different means: vFBAS by dynamic leader election [17]and FBA-SDN by offline transaction dissemination via nodes chosen at random [18]. In applied contexts, platforms like "Blockchain-based Guarantees of Origin Issuing Platform"[19] show how consensus reliability impacts real-world energy tracking systems. Other works like "Secure Architectures Implementing Trusted Coalitions for Blockchained Distributed Learning (TCLearn)" [20]and "Mitigation of Fake Data Content Poisoning Attacks in NDN via Blockchain" [21] reinforce the importance of robust consensus in adversarial environments. Formal verification efforts such as "Formal Modeling and Verification of a Federated Byzantine Agreement Algorithm for Blockchain Platforms" [9] and key infrastructure mechanisms like "Federated Distributed Key Generation for Blockchain-Based Federated Learning" [22] further provide technical depth on secure, decentralized coordination. Besides such practical implementations, formal frameworks like PBQS (Personal Byzantine Quorum Systems) as well as quantitative studies of splitting and blocking sets have contributed to a deeper theoretical understanding of safety and liveness in FBA networks [23].

The majority of FBA models still depend on static quorum-slice structures and do not react to altered network conditions or changes in node reputation. The other solve scalability, security, and decentralization with Hybrid Consensus[24]. Formal verification has been instrumental in ensuring safety under perfect conditions, but these guarantees do not imply runtime adaptability, especially when the system is confronted with real problems such as node failures or attacks [25], [26]. The risk is underscored by the experiment. If around 9% of the most influential validators are offline or compromised, the network's global consensus can be disrupted instantly. When the count of missing nodes gets to about 20%, quorum intersection may even fail completely. The reason for this fragility is that the influence in the FBA network is not equally spread; a small circle of validators eventually has most of the vital links, thereby making the topology susceptible to destabilization. Tumas et al. tried to lessen this weakness by changing the node links and rebuilding the network overlay. Their method increased the stability to around 45% [27] . However, the redesign is still static and thus unable to respond to validator behavior changes in real-time.

Studies on alternative consensus mechanisms reveal that reputation-based methods have started to become influential. For instance, EigenTrust has been considered in PBFT to identify disregarded replicas that are not trustable [28], [29]. Their method limits validator engagement through past trust measures, thus lowering the consensus communication that is necessary and increasing security in a very confidential type of environment [30]. Likewise, the same procedure is implemented in DPoS to get more precise validator rankings and to decrease the influence of unreliable participants [31]. These findings reveal that the evaluation of behavior can make a consensus protocol more robust. Nonetheless, no paper has so far considered the usage of adaptative reputation mechanisms in FBA, although the architecture of FBA is inherently compatible with dynamic trust evaluation. The majority of research works concerning FBA still depend on traditional methods. They usually apply static safety indicators or merely perform one-time changes of the network structure. These approaches are incapable of responding to changes in validator behavior or the occurrence of Byzantine challenges.

There have been upgrades to PBFT and DPoS through EigenTrust-based systems, yet, none of these endeavors have linked adaptive reputation scoring with the regeneration of quorum slices in a Federated Byzantine Agreement (FBA) network. The reason why this omission is significant is that FBA necessitates a mechanism which is able to uphold quorum connectivity instant by instant, particularly in an open and rapidly changing environment.

Consequently, this document puts forward and tests an Adaptive FBA framework. The layout comprises the validator reputation gotten from EigenTrust, temporal analysis through a sliding-window model, and the automatic reconfiguration of quorum. By the use of dynamic or adversarial conditions, among others, its objective is to secure FBA networks, keep consensus going and, ensure scalability is there in FBA nets, most especially when functioning under changing or hostile scenarios.

**2.2 Federated Byzantine Agreement Theory.**

David Mazières came up with the FBA in 2015 as part of the SCP [12]. It is a distributed consensus model. FBA lets a network of independent nodes come to an agreement on the state of a distributed ledger without having to coordinate with a central authority. Classical Byzantine Fault Tolerance (BFT) models need all nodes to agree on the same thing [14]. FBA, on the other hand, lets each node choose a set of trusted nodes, called a quorum slice, based on trust assumptions that are specific to that node.

FBA is structured around two fundamental concepts: quorum slices and quorum intersection. A quorum slice $S_v$ is a subset of nodes that a given node $v \in V$ considers sufficient to convince it of consensus. A quorum $Q \subseteq V$ is a set of nodes such that for every $v \in Q$, there exists a quorum slice $S_v \subseteq Q$ satisfying:

$$\forall v \in Q : \exists\ S_v \subseteq Q\ such\ that\ S_v \in QuorumSlice\ (V) \tag{1}$$

To ensure liveness, FBA requires that all quorums intersect. Formally, for any two quorums $Q_1, Q_2 \subseteq V$, the following condition must hold:

$$\forall Q_1, Q_2 \in Q : Q_1 \cap Q_2 \neq \emptyset \tag{2}$$

Quorum intersection in FBA systems makes sure that all quorums have at least one node in common. This stops agreements from conflicting and keeps everyone safe. The network must keep reaching consensus even when things go wrong in order to stay alive. Strongly Connected Components (SCCs) can be used to detect quorum intersections. a set $C \in SCC(G)$ satisfies $\forall\ u, v \in C: \exists$ path $u \to v$ and $v \to u$, ensuring mutual reachability and consistent consensus paths.

FBA shares these fault tolerance limits with classical byzantine systems, which it inherits from this structural base. The network structure and the configuration of the quorum must be able to support up to 2f+1 Byzantine nodes in order to be capable of maintaining quorum intersection and a sufficiently large honest majority thereby ensuring that these conditions are still fulfilled in the case of failures.

$$n \geq 3f + 1 \Rightarrow f \leq \left\lceil \frac{n-1}{3} \right\rceil \tag{3}$$

If the system is to remain functional, it can only allow up to 1/5 of the Byzantine nodes to be of a total number of the nodes. In case this threshold is exceeded, the property of the quorum intersection may be violated which, in fact, safety as well as liveness guarantees would not hold anymore[13]. Therefore, the issue of safety in FBA-based systems and their stability, particularly in the presence of adversaries in open environments, is largely dependent on the extent of the quorum overlap that is being maintained.

## 3. METHODOLOGY

**3.1** Overview of Experimental Design

This project implements a simulation-driven approach to verify the behavior of the Adaptive Federated Byzantine Agreement (FBA) under heavy network stress. Rather than depending on static models, the experiments bring back the real-world scenarios - validator churn, message latency, and even malicious behavior in some cases - to witness if the system can still save agreement and get back to normal by itself.

Each trial was performed in a designed simulation setting to ascertain the result was not by chance and could be reproduced. Validator nodes were structured into different organizations, and their efficiency was changed incrementally through various runs. In some instances, validators were made slower or disconnected; in other scenarios, quorum slices were broken to simulate structural faults. The network's ability to combine safety with liveness and also reconfigure its quorum structure utilizing adaptive reputation scores and fallback recovery logic was the thing to be measured.

The process is shown in Figure 1. First, a transaction comes into the network, then the Adaptive FBA module through the federated validators manages the validation process. Their quorum slices are continued to be updated dynamically from three main sources—reputation metrics, organizational affinity, and topological proximity. After consensus is achieved, the verified transaction is the ledger's next entry, thus the validation cycle is complete.

The model has fault tolerance as well. When a validator ceases to function or is marked as a malicious one, the adaptive control layer changes the quorum connections to continue agreement without the system stopping. All the simulations were carried out in Python, they combined real-time validator data from Stellarbeat.io with over 100 SCP log files collected in May and June 2025. The testing was done on an m2 macOS environment within Docker containers to simulate peer-to-peer interaction, and the resultant network graphs were made with the help of Gephi to investigate the changing connectivity patterns and trust evolution.

During the experiments, the following main performance indicators were monitored:
- Success rate in maintaining network connectivity, evaluated through Strongly Connected Component (SCC) detection;
- Frequency and structural properties of quorum slice regeneration events;
- Influence of validator reputation degradation on network cohesion;
- Failure tolerance concerning validator count, organization size, and trust threshold values.

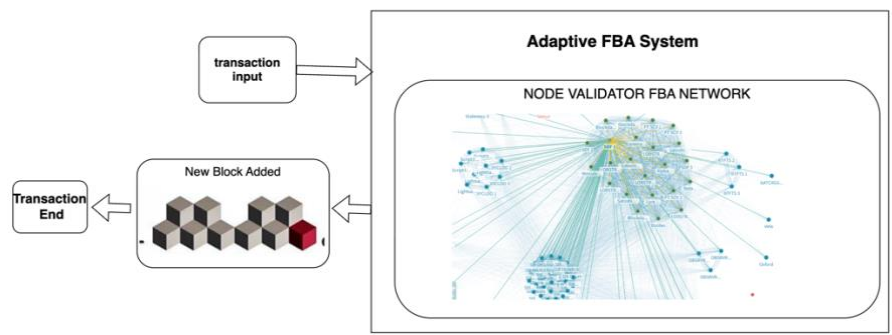

Figure 1. Adaptive FBA System Execution Flow

The experimental method here allows a quantitative and reproducing-level-of-the-same-experiment assessment of how well the Adaptive FBA system can sustain consensus when faced with a hostile environment. Furthermore, it sets up a comparative benchmark against classical FBA implementations, showing the model's capability of dynamically adapting while maintaining federated integrity.

3.2. System Architecture Overview

Figure 2 gives a summary of the Adaptive Federated Byzantine Agreement (FBA) setup. The structure is layers that are connected with each other and interact to ensure consensus is still reliable even when the network is in bad condition or some nodes behave unpredictably. The components number three:
1. Node Reputation Assessment Module
2. Quorum Slice Regeneration Module
3. Core Fallback Consensus Module

The process is getting started by collecting the validator data from Stellarbeat.io which includes the uptime record of each node, past validation history, quorum set, and general performance indicators. All these factors are fed to the reputation engine which decides how and when to form or adjust the quorum slices during operation.

Reputation evaluation is performed on two different but supplementary time scales.

Short-term a sliding-window monitor keeps an eye on each validator's decisions for the most recent consensus rounds. For every round i, the system stores whether the validator's vote was in line with the final network result. The combined outcome gives a live reputation score $R(j,t)$. Validators on the basis of this score are put in four different trust categories: Trusted, Semi-trusted, Cooldown, or Blacklisted.

Long-term a baseline EigenTrust is an overview of a node's past behavior-its uptime, quorum participation, and overall validation accuracy. This record is updated periodically (e.g., once a month). If a validator temporarily goes down to the Blacklisted category, the system checks its long-term EigenTrust value before it decides to allow it back to slice formation or not.

After these two evaluations are merged, the system remains sensitive to short-term network changes and at the same time it keeps the trust stable in the long term. Actually, reputation transitions are the reasons behind the regeneration of quorum slices in the given framework, which guarantees that the trust relationships will be always in line with the real performance of nodes. If a node's short-term behavior gets better, its previous EigenTrust background will help to fairly re-integrate it into the consensus pool.

The Quorum Slice Regeneration Module is on the spot when there is a disruption. The two main triggers for this process are:

(1) a structural failure, when the network loses global connectivity (this is the case when strongly connected component condition, SCC $\neq 1$, is violated); or

(2) a degradation of trust event, where based on sliding-window performance one or more validators being in the Blacklisted category.

Regeneration works to bring back quorum intersection by gradually reconstructing quorum slices, mainly counting on those nodes that have high reputation scores and at the same time are diverse in terms of organizations hence systemic bias is reduced. The module through its repeated rounds tries not only to get back to full network cohesion but also to keep liveness going. In case regeneration tries fail to recover connectivity, the system moves to the Core Fallback Module. This module employs a non-hierarchical, reputation-agnostic strategy where all validators have equal rights. Quorum slices result from uniform random sampling among possible nodes thus ensuring that basic consensus functionality is kept even when there is a heavy degradation or partial network collapse.

By means of this adaptive and layered design, the system is highly capable of dealing with validator churn, Sybil-type intrusions, and quorum fragmentation. Its feature of being able to change the consensus topology on the fly is what allows it to keep both safety and operational continuity going in a decentralized setting which is the basis for fault-tolerant blockchain infrastructures.

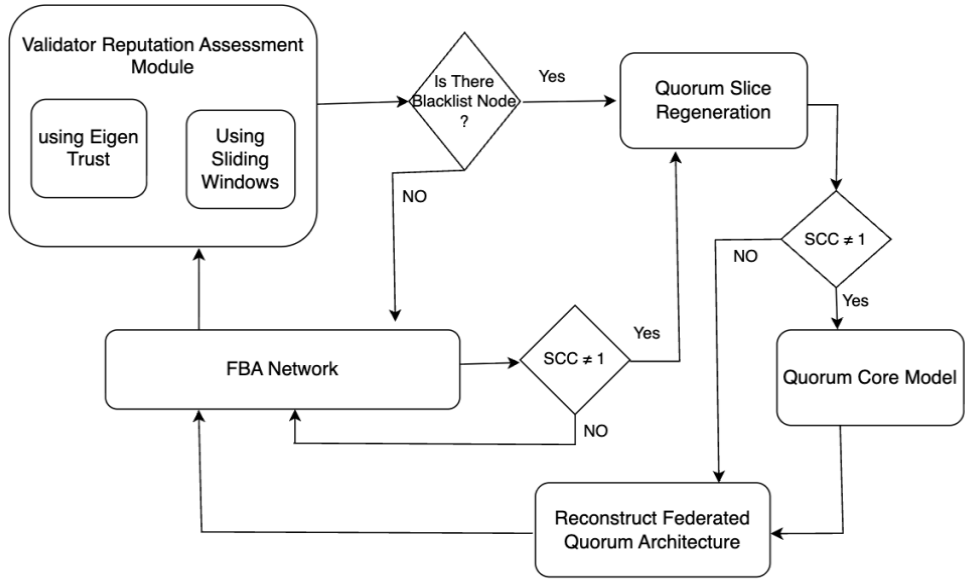

Figure 2. System Architecture Overview

### 3.3     Node Reputation Assessment Module

The Intelegent adaptive FBA model uses the EigenTrust algorithm and sliding window temporal analysis to create a framework for assessing node reputation dynamically. This system checks the reliability of validators in real time, allowing the quorum slice to change based on observations. Using the method suggested by Nugroho [32], we look at three main indicators to evaluate each validator node:

- isValidating: a node is currently taking part in the consensus protocol.
- fullValidator: a node open information to the other quorum.
- The overloaded : flag indicates when the resources are at full capacity.

Along with these operational metrics, behavioral assessments are based on how well a node's behavior matches quorum outcomes over time. In each round of consensus $i$, node $j$ receives a binary value denoted as $w_i^{(j)}$ according to the following criteria:

$$w_i^{(j)} = \begin{cases} 1, \text{if node j externalizes the majority quorum value} \\ 0, \text{ otherwise} \end{cases} \quad (4)$$

The reputation score of node j at time t is computed as the average over the sliding window:

$$R^{(j)}(t) = \frac{1}{N} \sum_{i=t-N+1}^{t} w_i^{(j)} \quad (5)$$

Based on this score, nodes are classified into four trust categories:
**Trusted** : $R^{(t)}(t) \geq \theta_1$,
**Semi-trusted**: $\theta_2 \leq R^{(t)}(t) < \theta_1$
**Cooldown**: $\theta_3 \leq R^{(t)}(t) < \theta_2$,
**Blacklisted**: $R^{(t)}(t) < \theta_3$, If the node doesn't agree with the quorum choice for five rounds in a row.

with thresholds set as $\theta_1 = 0.85$, $\theta_2 = 0.70$, dan $\theta_3 = 0.30$, based on empirically validated values as used in previous studies [33], where blacklisted nodes are treated as inactive or dead within the system.

### 3.4 Regeneration Quorum Module

To maintain system consistency and liveness under network dynamics, this module implements an adaptive regeneration process for quorum slices. When node failures or quorum intersection breakdowns occur, the system attempts to iteratively reconstruct valid slice configurations that preserve both structural integrity and quorum intersection.

Let $V = \{v_1, v_2, ..., v_n\}$ be the set of all validator nodes, where each node $v_i$ has a reputation score $R(v_i)$ and an organizational affiliation org$(v_i)$.

Each node attempts to construct a quorum slice $Q(v_i)$ that satisfies three validity criteria:
1. Average reputation $\tilde{R}(Q(v_i)) \geq r_{avg}$
2. Quorum Slice size based on organization.
3. Member of organisation

A quorum graph $G = (A, E)$ is then constructed, with directed edge $(v_i, v_j) \in A$ if and only if $v_j \in Q(v_i)$. The network achieves global quorum intersection if the graph has exactly SCC = 1.

### 3.4. Quorum Core Module

In scenarios where all attempts to establish a valid and strongly connected quorum structure fail, the system adopts a simplified Quorum Core strategy. This model disregards both organizational structures and node reputation filtering. Given a validator node set $V = \{v_1, v_2, ... v_n\}$ with $n = total\_nodes \in N$ }, all nodes are treated as peers and participate equally in a uniformly random slice formation mechanism.

Each node $v_i$ randomly selects s distinct nodes from $V \setminus \{v_i\}$ to construct its quorum slice, as defined by:

$$Q(v_i) = RandomSample(V \setminus \{v_i\}, s) \quad (6)$$

The quorum construction function is defined as:

$$Q: V \to P(V \setminus \{v_i\}) \text{ with } \lceil Q(v_i) \rceil = s \quad (7)$$

### 3.3. Experimental Setup

#### 3.3.1 Simulation-Based Experiments

An extensive of simulation experiments under controlled conditions mirroring Stellar-based networks validator behavior was conducted to assess the Intelegent Adaptive FBA framework. The simulation environment, which was coded with Python 3.11 and run on a macOS M2 system with 16 GB RAM, made use of several libraries. These libraries were networkx for quorum modeling, pandas for data handling, and matplotlib for visual analysis. Essential parts of the experiment such as slice generation and reputation evaluation were not only thought-out but also realized as separate functions to facilitate cross-configurations multiple experimental iterations. Validator data came from: (i) the actual datasets fetched through Stellarbeat.io (May–June 2025), and covering quorum sets, validator status, uptime, and organizational details; and (ii) a collection of 100 simulated SCP logs for computing sliding-window reputation metrics.

Each validator node carried a time-varying reputation score $R(v_i)$ that reflected their recent consensus behavior, while the EigenTrust scores functioned as a reference reputation and were updated on a monthly basis.

Experiment factors consisted of the number of validators in a quorum slice (from 3 to 7), the organizational diversity requirement (at least three different organizations per slice), and the cut-off points for average slice reputation. Each configuration was tried within 100 regeneration attempts at the most. If reconnecting the quorum was unsuccessful, a fallback core model was used. A quorum setting was considered correct only if it constituted one strongly connected component (SCC = 1). The regeneration events were triggered when SCC ≠ 1 or validator nodes were blacklisted.

Besides, in the same conditions, comparative simulations were run against conventional consensus protocols e.g. Proof-of-Work (PoW), Proof-of-Stake (PoS), Proof-of-Authority (PoA), PBFT, and classical FBA. The evaluation metrics included SCC success rate, regeneration frequency, failure tolerance, and system resilience.

In addition to the simulations, a Docker-based emulation platform was used to measure the performance overhead in the real world and confirm the robustness of the system. Validator nodes for PoW, PoS, and PoA were created using Hyperledger Besu and Geth, while PBFT nodes were using CometBFT (Tendermint). Both traditional and Intelegent Adaptive FBA versions were implemented as Python microservices inside the Docker network. One unified workload generator would issue transaction loads between 250 and 2,000 TPS, and a Prometheus-Grafana stack was used for throughput and latency monitoring.

Network delays that could be expected in the real world (60–120 ms with jitter) were introduced via tc/netem, and faults were injected by terminating validator containers or marking them as blacklisted within Adaptive FBA. The metrics gathered were throughput, finality latency, failure resilience, fork/orphan occurrence (for chain-based protocols), SCC connectivity (for FBA models), and regeneration events specific to Adaptive FBA. This hybrid approach—broad simulation combined with real-system emulation—allowed for an in-depth exploration of the parameter space and practical confirmation of the Intelegent Adaptive FBA's viability in operation.

## 4. RESULTS AND DISCUSSION

At the outset, we examined the intelegent adaptive FBA system under various real-world conditions. Besides that, in order to ascertain this, we conducted numerous experiments with various types of networks and kept track of the communication between the system's parts. We also undertook controlled experiments to understand how the system would behave in different environments. These experiments revolved around three parameters: the lowest average reputation within a quorum slice, the number of different organizations that the quorum slice is composed of, and the average size of the organizations from which members are coming. Moreover, the research separated these parameters and analyzed them deeply to find out how slight changes could have an impact on the system's total resilience, especially in the case of validator dropouts and network expansions.

The average reputation impact was the first parameter to be tested by us. It was implied by the experiment that the shift in the threshold immediately caused vast changes in the way locally connected quorum slices could be formed, and the trend shown in Figure 3 is a strong confirmation of that. The network was able to maintain global connectivity (100%) as long as the reputations threshold was less than 58. Nevertheless, the rate of forming SCCs significantly declined, after the threshold went beyond that point. This means that the point at which the average slice reputation is around 58 serves as a very critical cut-off. When the reputation threshold is regulated to not go beyond that value, the quorum slices remain overlapping, which is beneficial for the preservation of global quorum intersection as well as for the network's liveness.

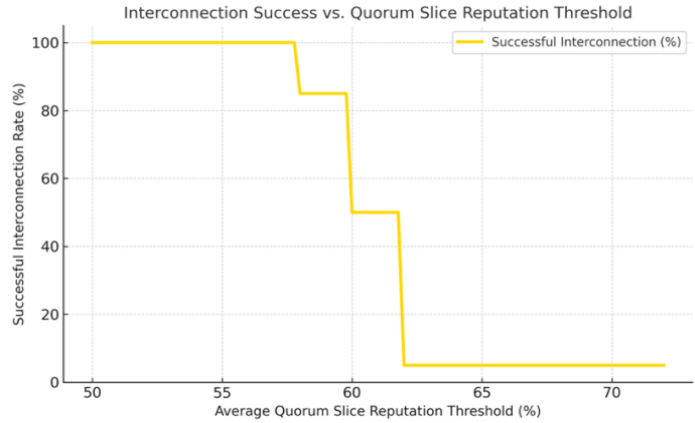

Figure 3. The average success rate in forming globally interconnected quorum slices

To ascertain the influence of diversity of organizations in each slice on network connectivity, we moved on to the next paragraph after that. A one-organization-only slice, for example, invariably is unable to overlap with others and the network finally disintegrates.. Just when a slice has two or more organizations, global connectivity can be seen starting and this can be kept up to an estimated number of ten organizations. To quench centralization, a quorum slice needs at least three different organizations to be involved, even though there are usually around seven in the slices of SCP deployments.

According to 4, the size of an organization influences how long the slices remain interconnected well. One- or two-member organizations tend to make unstable slices and may trigger disconnections at times. When an organization has at least three members, it is already somewhat stable. In reality, the highest level of connectivity is usually observed when an organization comprises four to six members. After that, there is a possibility of one organization dominating the slice which can lower the quorum diversity. Our next evaluation part was to analyze the system's behavior upon failure of validator nodes or when some of them are classified as unsafe, thus a scenario that hampers consensus in real networks.

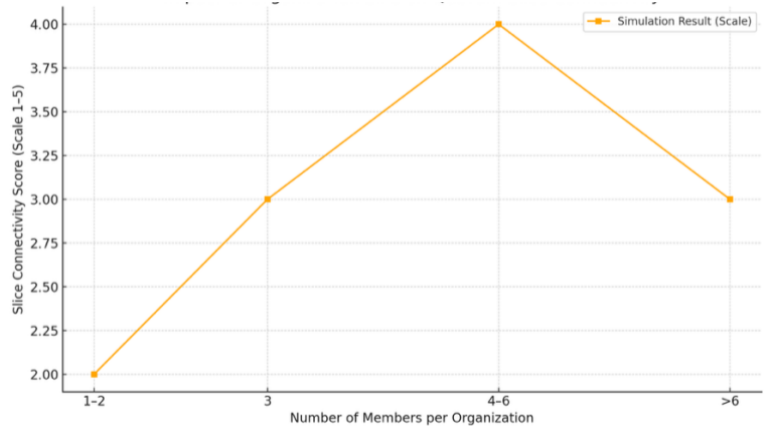

Fig. 4. Impact of Organization Size on Quorum Slice Connectivity.

The network simulations point to the network being able to sustain full quorum connectivity if the number of active validators is beyond 26. If the number goes down to 26 or less, the network also loses its quorum graph and is incapable of establishing global connectivity—the absence of which can be easily seen in Figure 5. This discovery is indicative of a very important validator availability threshold; the generative quorum-slice model is not able to assure quorum intersection once this point is crossed. To overcome the problem of connectivity in cases of low participation, the system has a Core Quorum mechanism.

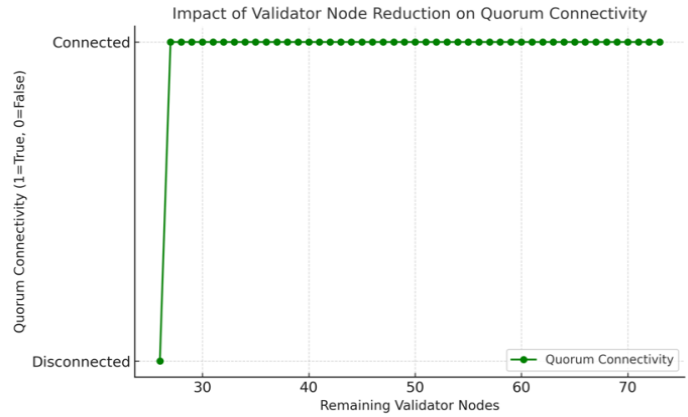

Figure 5. indicates a critical threshold of 26 validators;

Here, the limitations of an organization and the reputation factor are warranted, but each validator by randomly picking several peers—usually between three and five nodes—builds its slice. This method ensures that the quorum structure is still very tightly knit even if there is a severe splitting of the network. Scalability experiments involved the incremental growth both in the number of organizations and member size of each. The adaptive model does not lose global connectivity even when the number of organizations keeps increasing. The network is still sound, and quorum intersections are preserved even when there are one hundred validator nodes.

As seen in Table 1, the system is constantly creating slices and maintaining connectivity amid gradual organizational growth. In general, the findings of this study indicate the adaptive quorum-slice model being capable of network expansion while still ensuring the network remains live - given that organizational growth and slice diversity are developing at a similar speed.Table 1. Impact of Organization Growth on Quorum Connectivity

**Tabe 1 . Impact of Organization Growth on Quorum Connectivity**

| Added_Org_Members | Total_Organizations | Total_Nodes | Connected | Slices_Formed |
|---|---|---|---|---|
| 3 | 24 | 74 | TRUE | 74 |
| 5 | 25 | 79 | TRUE | 79 |
| 7 | 26 | 86 | TRUE | 86 |
| 9 | 27 | 95 | TRUE | 95 |
| 11 | 28 | 106 | TRUE | 106 |
| 13 | 29 | 119 | TRUE | 119 |
| 15 | 30 | 134 | TRUE | 134 |

Following the robustness assessment of Adaptive FBA, we put it side by side with several famous blockchain consensus protocols such as PoW, PoS, PoA, PBFT, and the classical version of FBA. Our comparison is not just about throughput or scalability but also about the trust each protocol model assumes, the way it finalizes the state changes, and its strength in the presence of faults or Byzantine actors. From this perspective, Adaptive FBA exhibits a nice and convincing balance decentralization, structural flexibility, and operational efficiency - a balance that both many typical blockchains and older FBA models have failed to achieve.

Table 2 shows the brief summary of different consensus mechanisms that are used in various blockchain systems. The mechanisms cover PoW, PoS, PoA, PBFT, classical FBA, and the study's Adaptive FBA. Along with the pure performance numbers, the comparison also considers how each protocol deals with trust, finality, scalability, and tough or adverse situations. PoW can be taken as a starting example.It is secure to an extreme degree as the breaking of the system will require enormous computational power but at the same time, there are grave energy consumption issues and throughput is very limited. While in PoS efficiency is raised as trust is given to token holders rather than computational power, the protocol carries the risk of structural failure: basically, the validators that have more resources will get more influence and thus the decentralized nature of the network can be compromised. This is a concern that keeps popping up. PoA, however, is based on another idea. As validators are already known and approved, the system turns out to be very fast and trustworthy, though trust is given to a handful of authorities. PBFT brings to the table good safety guarantees and finality that is deterministic but has difficulty scaling. When the number of validators increases, the overhead for communication grows rapidly and the performance goes down. Classical FBA is capable of handling scalability better as it employs federated quorum slices which enable decentralization to be maintained throughout the network. Nevertheless, classical FBA also faces the challenge of being static: when the behavior of validators changes or network topology alters, the integrity of the quorum can become compromised. The problems of classical FBA are the ones that Adaptive FBA deals with. It upgrades the classical version by adding features such as instantaneous reputation monitoring, on-the-fly slice regeneration, and a fallback option that gets activated when the failure occurs. Instead of using a fixed set-up, it is dependent on what the network is and thus it finds a solution.

Tabel 2 **Comparative Features of Consensus Mechanisms**

| Consensus Model | Trust Assumption | Finality Type | Scalability | Strengths | Limitations |
|---|---|---|---|---|---|
| **PoW** | Majority of hash power honest | Probabilistic | Low | Proven security, widely deployed | Energy intensive, low efficiency |
| **PoS** | Majority of stake honest | Probabilistic | Medium–High | Energy-efficient, better scalability | Risk of plutocracy and stake centralization |
| **PoA** | Pre-approved authority validators | Deterministic | High (small validator set) | High throughput, low latency | Centralization, relies on trusted authorities |
| **PBFT** | $f < n/3$ Byzantine nodes | Deterministic | Poor (>100 nodes unfeasible) | Strong safety, immediate finality | $O(n^2)$ overhead, not scalable |
| **FBA (Classic)** | Federated quorum slices | Deterministic | Medium–High | Decentralized trust via federations | Slice fragmentation, static trust mapping |
| **Adaptive FBA** | Federated trust + adaptive reputation | Deterministic | High (dynamic resilience) | Self-healing via slice regeneration & fallback | Higher complexity, requires continuous monitoring |

In figure 6, we present how transaction throughput (TPS) improves along with the increase of the number of validators. PoW remains unchanged. PBFT's throughput decreases rapidly due to its intensive messaging requirements. PoS and PoA are better to some extent, but only up to a certain level. Classical FBA is still quite strong before it starts to drop due to the large network. Intelegent Adaptive FBA is different from the rest. Beyond 100 validators, the network throughput increases to more than 12,000 TPS. Such a breakthrough is attributable to the adaptability of the model: it removes the validators with bad reputations, modifies the structures for the quorums as the network changes, and during disruptions, it is in the fallback mode that it relies on. This is a combined effect of the data in Table 2 and Figure 6. The Intelegent FBA is not only running the race at the same level as the others, but it is actually ahead of them. What it does is it brings in the advantages that current mechanisms lack. It has the qualities of being robust like BFT-style protocols yet is very fast and flexible which are the features that modern blockchain networks require.

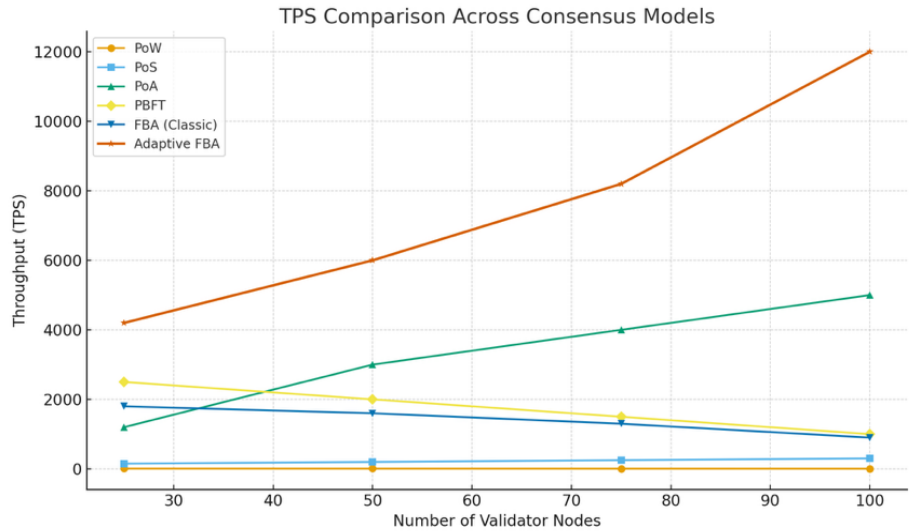

Figure 6 : comparasion TPS across Consensus Model

Although, significantly, it performs as if it had been purposely created for such scenarios in the actual world where the situations always change and cannot be predicted. The present work evidences that the adaptive FBA model is capable of maintaining network security and normal operation amid ever-changing conditions, such as validator failures and Byzantine behavior.

Intelegent adaptive FBA would be even more evident to the reader if the evaluation were only internal. PoW, PoS, PoA, and PBFT each have their own set of limitations despite being combined in various ways—scalability, decentralization, and fault tolerance cannot be fully balanced all at once. On the one hand, Classical FBA reports better scaling performance but usually breaks when heavy loads are put on the network. On the contrary, Adaptive FBA keeps stable quorum connections and is able to attain throughput several times higher than that of classical FBA. It almost achieves the performance of PoA and PBFT while at the same time not suffering from the problem of centralization which is characteristic of both. In fact, all of this evidence leads to a single inference: Adaptive FBA represents exactly the features that current consensus models lack and therefore occupies the central position among them. It possesses the qualities of being flexible, durable, and efficient which happen to complement each other perfectly in blockchain worlds that are volatile and unpredictable by nature.

## 5.  CONCLUSION

The presented paper approach dynamically reconfigures quorum-slice structures based on the reputation of nodes. This allows the network not only to maintain real-time connectivity but also to make the overall consensus process more reliable. Our tests reveal that the network is capable of locally identifying situations where the quorum is broken and consequently switching to a backup mode that enables the recovery of the global connection. The network keeps a valid quorum graph while being in Regeneration Quorum mode as long as 26 or more validators are still active. Theoretically speaking, the network is even able to carry out its operations with only three trustworthy validators in Core Quorum mode which is an extremely impressive level of robustness for a federated model. The system is also optimized when the average reputation threshold for the nodes is approximately 56.

The ways in which reputation is determined is at the moment a major drawback of the system—most of the time, this is done by counting the votes that are in line with the final decisions of the committee. This, however, opens up a risk: a node may act properly in most cases but still send out harmful messages when there is a crucial failure. Therefore, research work to come is expected to be committed to advanced security models that can dynamically enforce reputation and be enabled by real-time resource monitoring, cross-layer anomaly detection, and adaptive methods such as reinforcement learning (RL)..